\documentclass[aps,prl,amsmath,amssymb,nofootinbib,twocolumn]{revtex4-1}
\usepackage{graphicx}
\usepackage{dcolumn}
\usepackage[utf8]{inputenc}
\usepackage{tikz-feynman}
\usepackage{tikz}
\usepackage{feynmp} 
\usetikzlibrary{shapes.misc}
\usepackage{footmisc}

\usepackage{bm}
\usepackage{hyperref}
\usepackage{xcolor}
\usepackage{float}
\begin{document}
\setlength{\abovedisplayskip}{5pt}
\setlength{\belowdisplayskip}{5pt}
\setlength{\abovedisplayshortskip}{5pt}
\setlength{\belowdisplayshortskip}{5pt}

\preprint{}

\title{Anapole moment of neutrinos and radioactive sources near liquid xenon detectors}
\author{Gonzalo Herrera}
\author{Patrick Huber}
\affiliation{Center for Neutrino Physics, Department of Physics, Virginia Tech, Blacksburg, VA 24061, USA}

\begin{abstract}
We show that placing a radioactive source such as ${}^{51}$Cr near a liquid xenon detector may allow to detect the contribution induced by the anapole moment to neutrino-electron scattering in the Standard Model at the 1-2 $\sigma$ level. Although the anapole moment of neutrinos induces a scattering rate with the same spectral shape as the neutral and charged current contributions, exposures of $\sim$ 60 tonne $\times$ source run at XENONnT or XLZD may be enough to accumulate sufficient statistics for a detection. We also discuss a simple model where the anapole moment of neutrinos is enhanced or decreased with respect to the SM expectation, further demonstrating how a potential measurement of the anapole moment of neutrinos would allow to constrain new physics.
\end{abstract}

\maketitle

\section{Introduction}

Although neutrinos are electrically neutral in the Standard Model (SM), the Lorentz and gauge symmetries permit a magnetic, electric and anapole moment at one-loop \cite{Nieves, Kayser, Giunti:2014ixa}. Multiple studies have been dedicated to investigate the phenomenological consequences of the magnetic and electric moment of neutrinos, \textit{E.g }\cite{Cisneros:1970nq,Barbiellini:1987zz, Grimus:1997aa, Raffelt:1999gv,Studenikin:2013my,Chen:2014dsa, Canas:2015yoa, Hsieh:2019hug, AristizabalSierra:2021fuc, Khan:2022bel, Khan:2022akj, AtzoriCorona:2022jeb, Coloma:2022umy, Coloma:2022avw, A:2022acy, Herrera:2023xun, AristizabalSierra:2022axl}. However, current probes are yet far from being able to probe the expected values in the SM, which are tiny due to their proportionality to the neutrino mass \cite{Fujikawa:1980yx, Shrock:1982sc}. The anapole moment was however not so widely explored. Proven to be a gauge invariant quantity, its expected value for neutrinos in the SM is well known \cite{Musolf:1990sa,Bernabeu:2000hf, Cabral-Rosetti:2002zyl, Bernabeu:2002nw, Fujikawa:2003ww}. Bounds on its value from reactor and solar neutrino experiments have been derived, lying remarkably close to the SM prediction, just roughly 1 to 2 orders of magnitude below depending on the experiment under consideration \cite{Hirsch:2002uv, TEXONO:2009knm, Cadeddu:2018dux,A:2022acy, AtzoriCorona:2022qrf, MammenAbraham:2023psg, Giunti:2023yha}.

We propose to place radioactive neutrino sources such as ${}^{51}$Cr near liquid xenon detectors to search for the neutrino anapole moment.  The monoenergetic neutrino flux  of  a MCi ${}^{51}$Cr source placed at 1-10 meters from the detector can exceed the solar neutrino flux by orders of magnitude, which may allow to detect electronic recoil rates sensitive to the neutrino anapole moment with current and near future planned experimental exposures at large volume liquid xenon experiments \cite{LZ:2022lsv, XENON:2023cxc}.

Furthermore, we discuss a model where neutrinos may acquire enhanced or suppressed anapole moments with respect to the SM, due to their one-loop mixing with a dark photon, under which new light dark sector particles are charged. In this manner, we show how a future measurement of the neutrino anapole moment may allow to constrain new physics.

\section{Neutrino-electron scattering rate in liquid xenon}
Diagonal neutrino interactions induced by an anapole moment arise from the Lagrangian

\begin{equation}
	\mathcal{L} =  \frac{a_{\alpha}}{2} \overline{\nu_{\alpha}} \gamma^\mu \gamma_5 \nu_{\alpha} \partial^\nu F_{\mu\nu},
\end{equation}

with $\alpha=e,\mu,\tau$. In the SM,

\begin{equation}
	{a}^{\rm SM}_{\alpha}\simeq\frac{G_F}{24\sqrt{2}\pi^2}\left(3-2\log\frac{m^2_\alpha}{m_W^2}\right)
\end{equation}

which yields values of the diagonal anapole moment of neutrinos of $a_{ee}^{\mathrm{SM}}=6.8 \times 10^{-34} \mathrm{~cm}^2$, $a_{\mu \mu}^{\mathrm{SM}}=4 \times 10^{-34} \mathrm{~cm}^2$, and $a_{\tau \tau}^{\mathrm{SM}}=2.5 \times 10^{-34} \mathrm{~cm}^2$ \cite{Cabral-Rosetti:2002zyl, Giunti:2014ixa}.

Neutrinos with flavor $\alpha$ predominantly interact with the electrons in the atom elastically $\nu_{\alpha} e^{-} \rightarrow$ $\nu_{\alpha} e^{-}$. We approximate the elastic differential scattering cross section with the electrons in the atom as
\begin{equation}
\frac{d \sigma_{\alpha}}{d T}=\sum_{k=1}^Z \theta\left(T-E_{\text {bind }}^a\right) \frac{d \sigma_{\nu_{\alpha} e_k^{-} \rightarrow \nu_{\alpha} e_k^{-}}^{\text {free }}}{d T},
\end{equation}
where $T$ is the recoiling kinetic energy of the electron, $\frac{d \sigma^{\rm free}}{d T}$ is the scattering cross-section with a free electron, and $E_{\text {bind }}^a$ is the binding energy of the electron $k=1 \ldots Z$, with $Z$ the atomic number of the atom.

The free neutrino-electron scattering cross section induced by the anapole moment can be parametrized by means of a redefinition of the weak mixing angle

\begin{equation}
\sin ^2 \theta_W \rightarrow \sin ^2 \theta_W\left(1-2 m_W^2 a_{\alpha}\right),
\end{equation}
and the diagonal electroweak neutrino-electron scattering cross section reads \cite{Vogel:1989iv}
\begin{equation}
\frac{d \sigma_{\alpha}}{d T}=\frac{2 G_F^2 m_e}{\pi}\left[g_{L,\alpha}^2+g_{R,\alpha}^2\left(1-\frac{T}{T_\nu}\right)^2-g_{L,\alpha} g_{R,\alpha} \frac{m_e T}{T_\nu^2}\right],
\end{equation}
where $g_{L,e}=\left(g_V+g_A\right) / 2+1$, $g_{L,\mu}=g_{L,\tau}=\left(g_V+g_A\right) / 2$, and $g_{R}=\left(g_V-g_A\right) / 2$, with  $g_V=-1 / 2+2 \sin ^2 \theta_W$, and $g_A=-1 / 2$. Now, we can calculate the differential recoil rate at liquid xenon detectors via
\begin{equation} \label{eq:recoil_rate}
\frac{d R}{d T^{\prime}}=N_T \mathcal{E} \sum_{\alpha} \int d T \int_{T_\nu^{\min }}^{T_\nu^{\max }} d T_\nu \left\langle \frac{d \phi^{\alpha}}{d T_\nu} \right\rangle \frac{d \sigma_{\alpha}}{d T} \epsilon\left(T^{\prime}\right) \lambda\left(T^{\prime}, T\right),
\end{equation}
where the total number of target atoms is $N_{T} = 6.02 \times 10^{26}/A$ per kg, with $A$ the atomic mass of the detector atom in atomic units, and $\mathcal{E}$ is the exposure in units of kg $\times$yr. $\epsilon(T^{'})$ is the detector efficiency in terms of the reconstructed energy $T^{'}$, and $\lambda(T^{'},T)$ is a normalized Gaussian smearing function to account for the detector energy resolution \cite{XENON:2023cxc,LZ:2022lsv,Khan:2020vaf}. The integration limits over recoil energy are determined by the energy threshold and maximum energy within the region of interest of the experiment. We point out that the sensitivity to the anapole moment of neutrinos could be significantly increased if these experiments extended their region of interest towards larger energies than currently, which is $\sim$ 70 keV. For instance, extending the region of interest to 140 keV, which was done for a low-exposure search of electron recoils in \cite{XENON:2022ltv}, we find an enhancement on the total number of anapole-induced recoil events of $\sim 1.8$. Finally, using the XENONnT efficiency function and energy resolution \cite{XENON:2023cxc}, we can calculate the total rate. Although we focus on the XENONnT experiment, our proposal could be applied in other large volume liquid xenon experiments like PandaX-4T or LUX-ZEPLIN \cite{PandaX-4T:2021bab, LZ:2022lsv}.
The minimum neutrino energy needed to induce a recoil of energy $T$ is given by
\begin{equation}
T_\nu^{\rm min}=\frac{T+\sqrt{2 m_e T+T^2}}{2},
\end{equation}
and $T^{\rm max}$ corresponds to the maximum neutrino energy from $\left\langle \frac{d \phi^{\alpha}}{d T_\nu} \right\rangle$, the average neutrino flux per flavor arriving to the detector for a given time exposure. We will be considering the pp-chain solar electron neutrino flux \cite{Vitagliano:2019yzm}, with survival probability of electron neutrinos of $P_{ee}=0.553$. Further, we will consider the neutrino flux stemming from a ${}^{51}$Cr source \cite{Peng:2022nvi,Coloma:2014hka}. The ${ }^{51}$Cr source produces electron neutrinos via the reaction ${ }^{51} \mathrm{Cr}+e^{-} \rightarrow { }^{51}\mathrm{V}+\nu_e$ with a half-life of 27.7 days. Given this half-life, in the following we will denote exposures in units of tonnes $\times$ source run, with a ${}^{51}$Cr source run corresponding to 2 months. The spectrum of electron neutrinos from a ${ }^{51} \mathrm{Cr}$ source consists of four monoenergetic lines with $0.427$ , $0.432,0.747$, and $0.752$ MeV energies, with branching ratios of $0.09,0.009,0.816$, and 0.085, respectively. We will be considering a ${ }^{51} \mathrm{Cr}$ source with an activity of 3 $\mathrm{MCi}$, and placed at 1 m from the detector, off-axis and at an intermediate height of the detector tank. We then simulate the average distance to the detector, using the dimensions of the XENONnT tank, with diameter at the base of 132.8 cm and height of 148.6 cm \cite{XENON:2024wpa}. In our set-up, we find an average distance of the source to the detector of 131.9 cm.  The shielding of the XENONnT detector in LNGS is thin enough to allow a source to be placed 1 m from it at an intermediate height. 

The measurement considered here can be done with either neutrinos or antineutrinos. The best available antineutrino source in the relevant energy range would be a nuclear reactor due to its extremely high antineutrino flux and low-energy spectrum. However, the detector needs to be deep underground to suppress cosmogenic backgrounds and there are no deep underground nuclear reactors. An intense beta decay radioactive source in the kCi range has been considered for Borexino~\cite{Borexino:2013xxa} but there was an accident during the separation of the isotope from spent nuclear fuel~\cite{soxpnas}. Neutrino sources at the keV--MeV scale are either beta-minus emitters, positron emitters or decay via electron capture. Electron capture  sources are attractive since very little non-neutrino radiation is emitted, for a pure electron capture decay only a few Auger electrons and x-rays are emitted. Additional radiation from decays into excited states can be present. This greatly facilitates radiation shielding and makes these sources overall much safer to handle. Also the resulting mono-energetic neutrino flux reduces systematic uncertainties significantly.  Electron capture sources using $^{51}$Cr~\cite{GALLEX:1994rym,Abdurashitov:1996dp} and $^{37}$Ar~\cite{Abdurashitov:2005tb} have been used to calibrate the solar neutrino experiments GALLEX and SAGE and more recently a $^{51}$Cr source has been used for the BEST experiment~\cite{Barinov:2021asz} to look for sterile neutrinos. Given this recent experience and the fact that $^{37}$Ar requires a fast neutron flux and hence is considered challenging to produce, we will focus on a $^{51}$Cr source; see also Refs.~\cite{Grieb:2006mp,Coloma:2014hka,Coloma:2020voz,Peng:2022nvi}. We assume the same parameters as for the BEST source as detailed in Ref.~\cite{Danshin:2022wiz}, specifically we use an activity of 3.1\,MCi. We also discuss how future liquid xenon experiments, with detector masses as those proposed in the future XENON/DARWIN and LUX-ZEPLIN collaboration (XLZD) \cite{Aalbers:2022dzr}, could receive a sizable contribution in the scattering rate due to neutrino anapole moments during the lifetime of a single source of ${}^{51}$Cr, corresponding to 27.7 days. 

\begin{figure}[t!]
    \includegraphics[width=\linewidth]{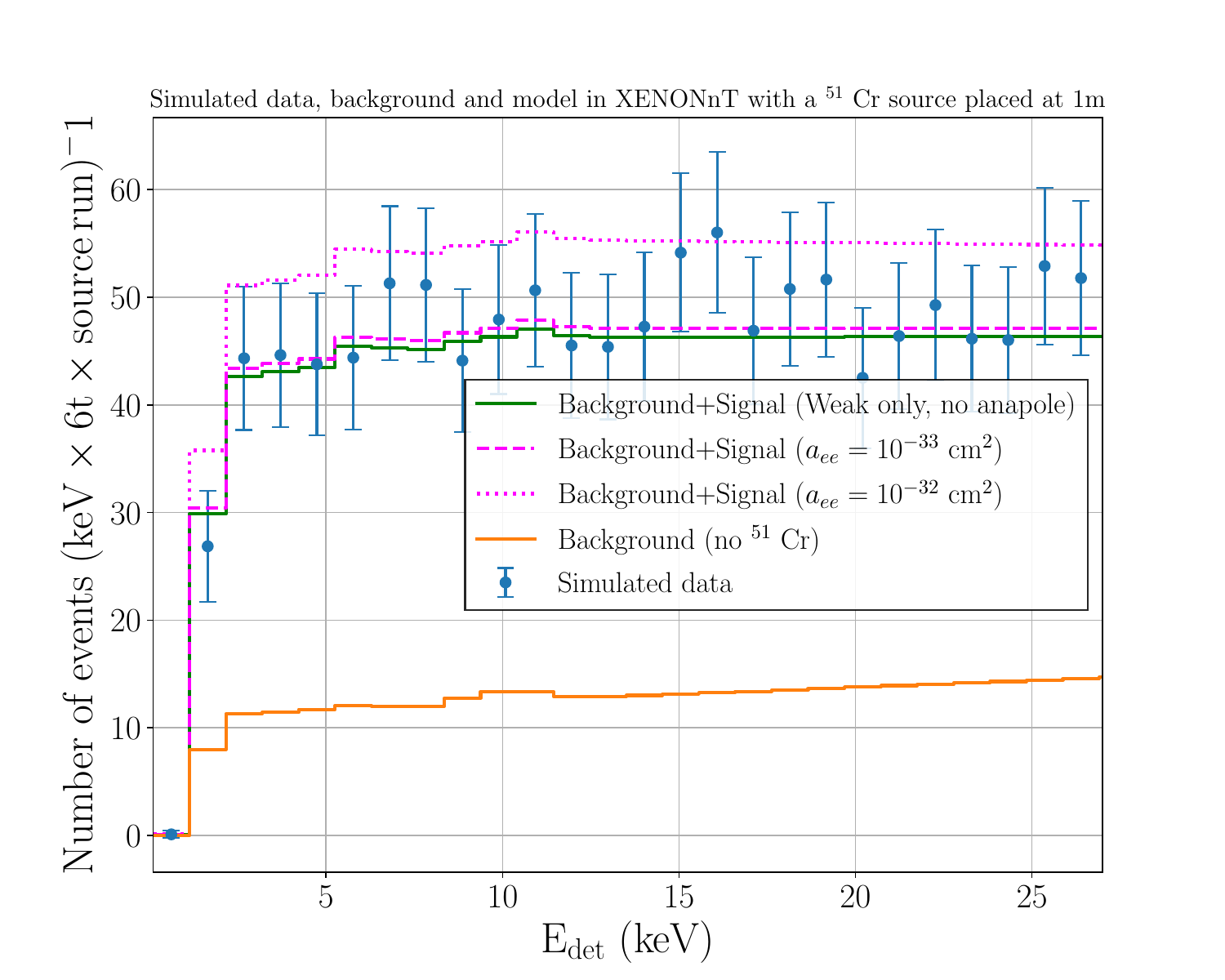}
    \caption{Simulated ionization rate in XENONnT, including data with systematic uncertainty (blue), background model without the Chromium source (orange), background plus weak interaction contribution with the Chromium source (green) and signal model for different values of the anapole moment (magenta) in XENONnT with a ${}^{51}$Cr source placed at an average distance of 131.9 cm from the detector (see main text for details).}
    \label{fig:binned_rate}
\end{figure}

\section{Sensitivity to the electron neutrino anapole moment}
First, we will simulate the binned data and background of the XENONnT experiment with a ${}^{51}$Cr source placed off-axis at 1m from the detector, including the weak interaction rate of neutrinos from the radioactive source. We will neglect additional backgrounds that may arise from the source impurities and gamma rays from ${}^{51}$Cr itself. It has been shown that both intrinsic gamma rays and  those from activated impurities can be effectively shielded by 2\,cm of tungsten combined with 40-45\,cm of depleted uranium~\cite{Huber:2022osv}. The simulated data from XENONnT is normalized to an exposure of 6 tonne $\times$ source run, equivalent to 1 tonne $\times$ yr. We use the background model at XENONnT from \cite{Aprile:2022vux}, with same energy threshold, and simulate the data from a Poissonian distribution of this background model plus the weak interaction rate from neutrinos from the Chromium source, fixing at this level of the analysis sin$^2 \theta_{W}=0.231$. We consider the statistical uncertainty of the simulated data only, implicitly assuming that the systematic uncertainties are very small. We perform this calculation for the process $\nu_{e}e \rightarrow \nu_{e}e$, via the SM neutral and charged current contributions and via the anapole moment. As expected, the anapole moment contribution presents the same spectral shape as the W/Z mediated contribution, which requires large statistics to differentiate both components within statistical errors. Due to the larger neutrino flux from ${}^{51}$Cr, we find that the recoil rate is enhanced with respect to the solar one by roughly a factor of $\sim 19.5$, and extends towards larger energies than the endpoint of the solar $pp$ neutrino flux. This yields sizable number of events induced from a neutrino anapole moment in the detector for $\sim$ 6 tonne $\times$ source run exposures, within current reach of XENONnT.

Figure \ref{fig:binned_rate} shows the simulated data and background model in XENONnT, including the contributions from the anapole moment of electron neutrinos with different values. As can be appreciated in the plot, for this exposure, the statistical uncertainty eclipses any differences in the ionization rates induced by the value of the anapole moment expected in the SM. However, if the anapole moment became larger due to Beyond the SM contributions, it could be detectable with $\sim$ 6 tonne $\times$ source run exposures. For instance, it can be appreciated in the Figure that an anapole moment of $a_{ee}=10^{-32}$ cm$^2$ would induce a rate that would exceed the statistical error bars at most energy bins. Such enhanced values of the anapole moment can be induced by a MeV scale millicharged sector that further couples at tree level to neutrinos. We will discuss such a model later in the manuscript. In the following, we demonstrate that with sufficiently large exposures, the statistical uncertainty may not mask the differences in the recoil rate induced by the neutrino anapole moment as expected in the SM, due to their scaling with exposure $\mathcal{E}$ as $\sigma \sim \sqrt{\mathcal{E}}$, while the rate induced by the anapole moment scales linearly $ N_{\rm exp} \sim \mathcal{E}$.

\begin{figure}[t!]
    \includegraphics[width=\linewidth]{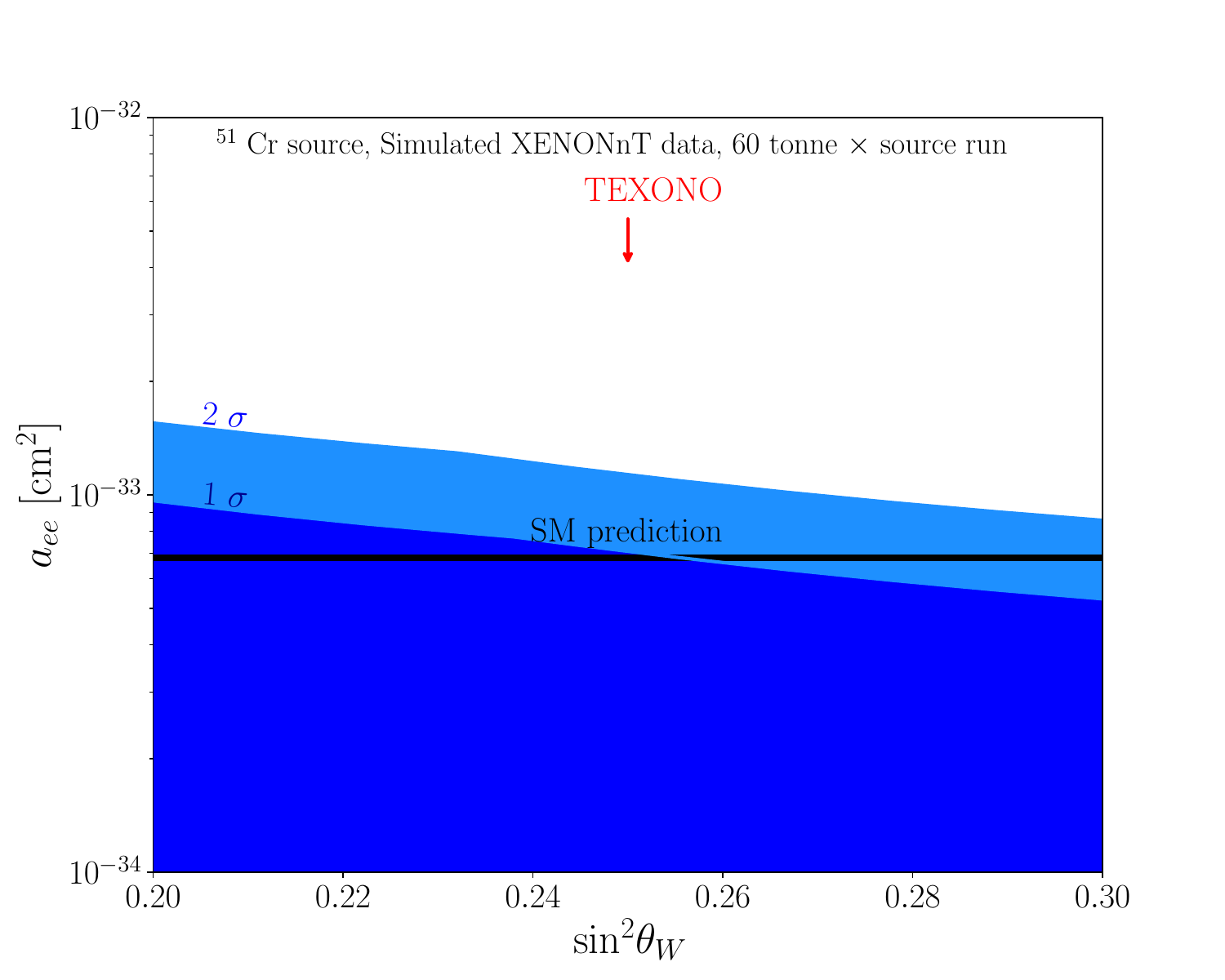}
    \caption{1$\sigma$ and 2$\sigma$ level contours on the electron neutrino anapole moment with XENONnT and a ${}^{51}$Cr source, and a exposure of 60 tonne $\times$ source run, equivalent to 10 tonne $\times$ year). The strongest limit on the neutrino anapole moment from TEXONO, $a_{ee} \leq 5.5 \times 10^{-33}$ cm$^2$, is shown as a red arrow \cite{TEXONO:2009knm}. For comparison, we show the SM prediction on the electron neutrino anapole moment as a black line.}
    \label{fig:limit_anapole_51Cr-electrons}
\end{figure}

We will derive projected constraints on the diagonal anapole moment of electrons neutrinos, $a_{ee}$, from a non observation of an excess of events in a simulation of a future XENONnT experiment exposed to a Chromium source. We will perform a reduced binned $\chi^2$ analysis in the region of interest of XENONnT, where the $\chi^2$ function is defined as
\begin{equation*}
\small{
\chi^{2}=\sum_{i}\frac{(N_{\rm obs}-N_{\rm bck}-N_{\rm exp}(a_{ee}, \mathrm{sin}^2\theta_{W}, u_{Cr}))^2}{\eta_i^2}}
\end{equation*}
and the index $i$ runs over all bins in the experiment, $N_{\rm obs}$ is the number of simulated observed events at XENONnT, $N_{\rm bck}$ is the background model without the ${}^{51}$Cr source, and $\eta$ is the statistical uncertainty on the observed number of events.$N_{\rm exp}(a_{ee}, \mathrm{sin}^2\theta_{W}, u_{Cr}, u_{m})$ is the number of expected events from neutrino-electron scattering induced by the flux of neutrinos produced at the Chromium source, which depends on the precise value of the weak mixing angle, which is uncertain, on the uncertainty on the activity of the Chromium source $u_{Cr}$, on the uncertainty on the fiducial mass of the liquid xenon detector, $u_{m}$, and on the electron neutrino anapole moment itself $a_{ee}$. In reality, the off-diagonal anapole moments of the electron neutrinos from the Chromium source, and the muonic and tauonic anapole moments of solar neutrinos would also induce scatterings in the detector. For the SM predicted values of the neutrino anapole moment, we find those contributions to be subdominant (at least $\sim 60 \%$ smaller than the diagonal contribution), so we will neglect them in the rest of the analysis \footnote{Our sensitivity prospects on the electron neutrino anapole moment for the Chromium source can also be interpreted in terms of a linear combination of the diagonal and off-diagonal contributions. The diagonal contribution largely dominates, since the off-diagonal contributions are suppressed by the PMS matrix entries via  $a_{\nu_{\ell^{\prime} \ell}}=\sum_{k, j} U_{\ell^{\prime} k} a_{\nu_{k j}} U_{\ell j}^*$, which amounts to a $\sim 50-60\%$ suppression w.r.t to the diagonal contribution. Our analysis focuses on probing solely the diagonal electron neutrino anapole moment, which is a conservative approach.}
\begin{figure*}[t!]
    \includegraphics[width=0.49\linewidth]{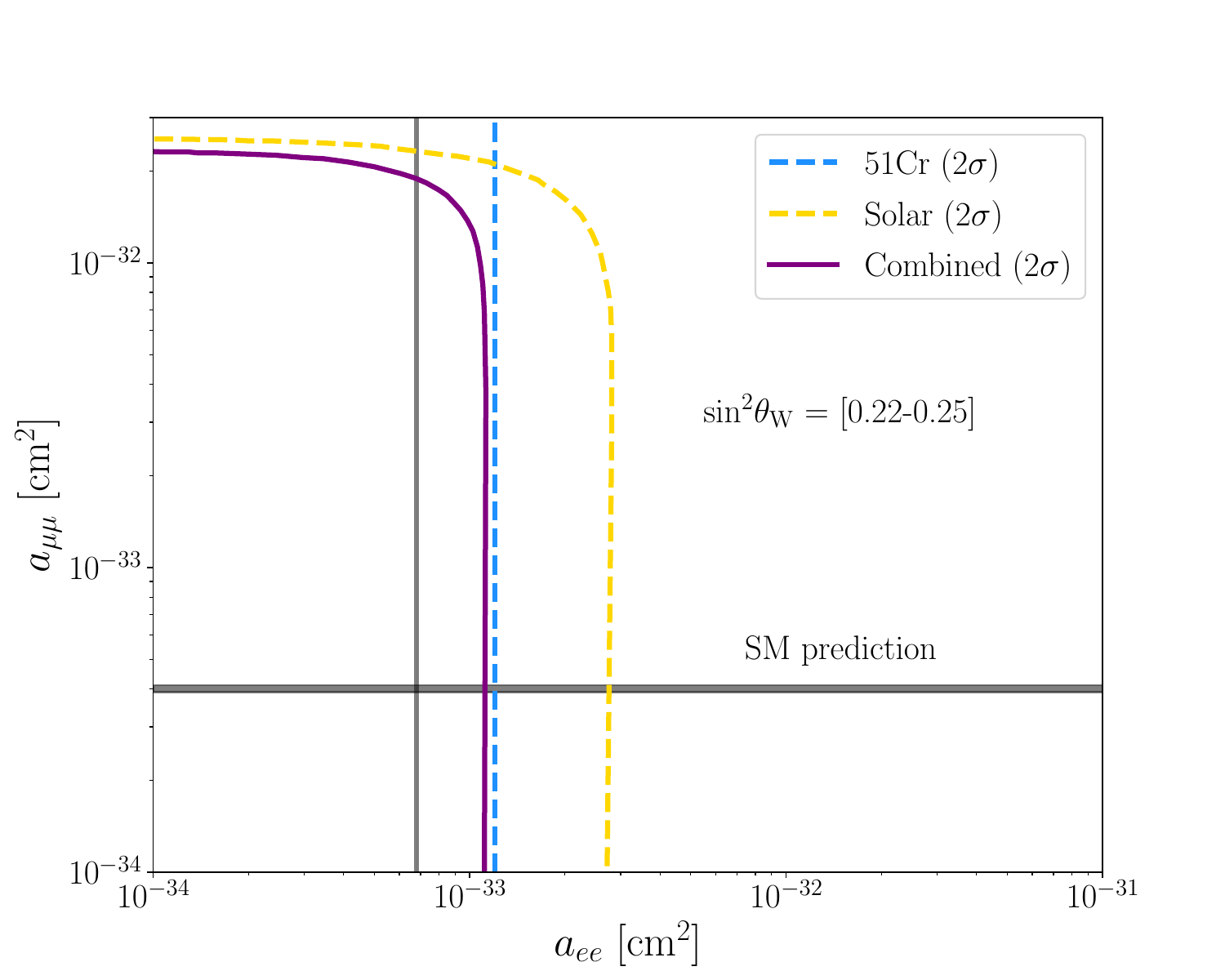}
    \includegraphics[width=0.49\linewidth]{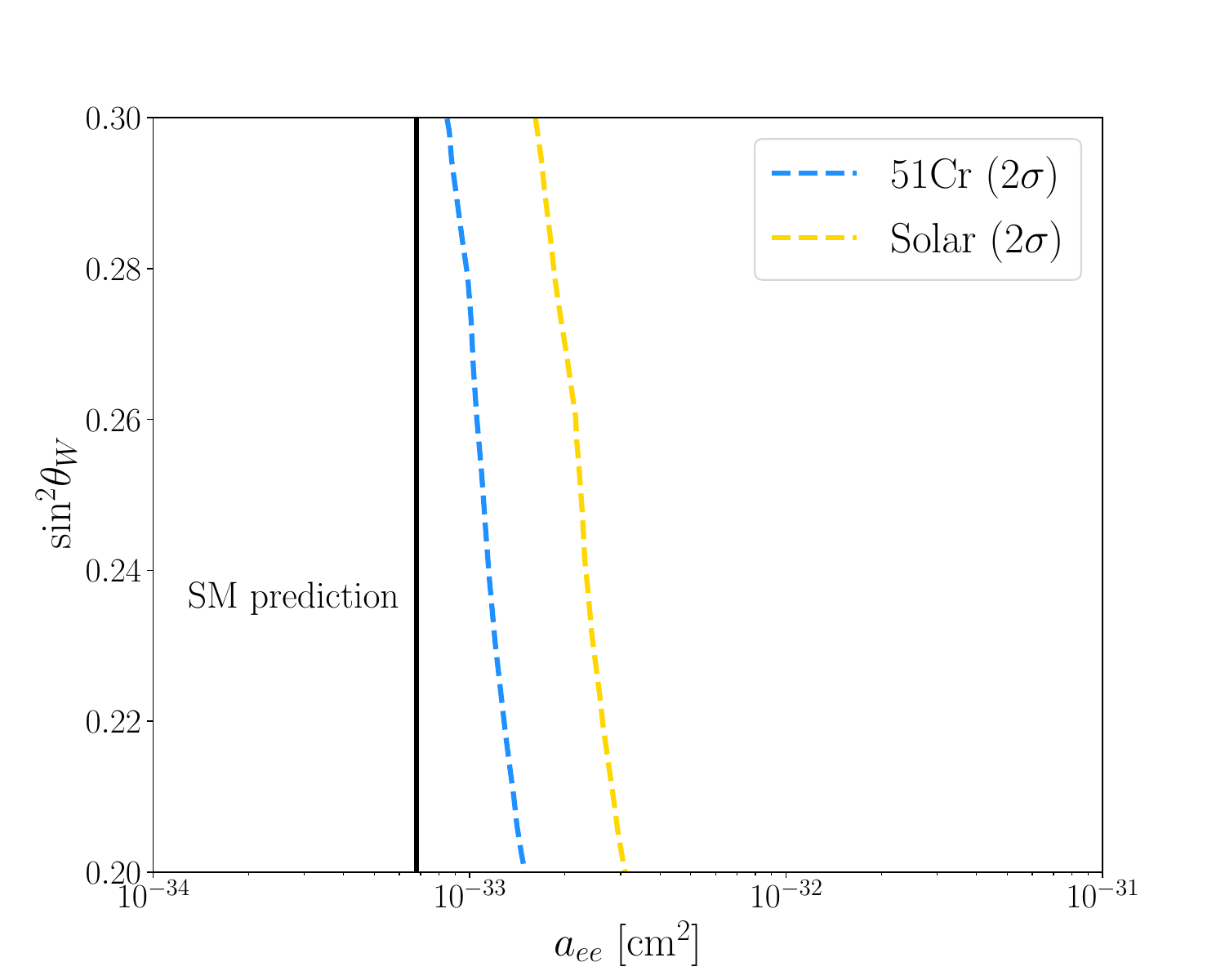}
    \caption{\textit{Left plot}:  2$\sigma$ sensitivity on the combination of electron neutrino anapole moment and muon neutrino anapole moment with XENONnT. The sensitivity is derived from the combination of a ${}^{51}$Cr source, and a exposure of 60 tonne $\times$ source-run, and a run without the Chromium source, but considering a solar neutrino flux with a exposure of 195 tonne $\times$ years, which would yield comparable rates to the Chromium source with 60 tonne $\times$ source run. The solid black lines indicate the SM predictions for the diagonal anapole moment of electron neutrinos and muon neutrinos. \textit{Right plot}: 2$\sigma$ sensitivity from the ${}^{51}$Cr source with an exposure of 60 tonne $\times$ source run (blue), and from solar neutrinos with an exposure of 195 tonne $\times$ years (orange), on the parameter space of electron neutrino anapole moments and the weak mixing angle. The SM predictions for the anapole moments are shown as vertical and horizontal lines.}
    \label{fig:anapole_mue}
\end{figure*}
We can derive an upper limit on the uncertain parameters of the $\chi^2$ function with the desired significance $\sigma$ and degrees of freedom $d.o.f$ as \cite{Wilks:1938dza}
\begin{equation}
\chi^2(a_{ee}, \mathrm{sin}^2\theta_{W}, u_{Cr}, u_{m}) - \chi^2_{\rm min} \leq \chi^{2}_{\sigma, \rm d.o.f}
\end{equation}
where $\chi^2_{\rm min}$ is the minimum of the function over all parameters, and $\chi^{2}_{\sigma, \rm d.o.f}$ is obtained from the statistical $\chi^2$ distribution. In practice, we will constrain only the anapole moment and weak mixing angle together, and we will minimize our $\chi^2$ over the uncertainty in the purity of the Chromium source, which we take to be 1$\%$, although it is reported by the BEST collaboration to be as small as 0.23$\%$ \cite{Danshin:2022wiz}, and over the uncertainty of the detector mass, which is 2.75$\%$ for a recent analysis at XENONnT \cite{XENON:2023cxc}. The uncertainty on the detector mass and the Chromium source, parametrized with $u_{m}$ and $u_{Cr}$, act as a correction on the normalization of the recoil rate. Concretely, we introduce the following modifications in Eq. \ref{eq:recoil_rate}
\begin{equation}
\mathcal{E} \rightarrow (1+u_m)\mathcal{E} 
\end{equation}
with $u_m \in (-0.0275,0.0275)$ for the detector mass uncertainty, and
\begin{equation}
\left\langle \frac{d \phi^{\alpha}}{d T_\nu} \right\rangle \rightarrow (1+u_{Cr})\left\langle \frac{d \phi^{\alpha}}{d T_\nu} \right\rangle 
\end{equation}
with $u_{Cr} \in (-0.01,0.01)$ for the uncertainty in the Chromium source activity.
In Figure \ref{fig:limit_anapole_51Cr-electrons}, we show the 1$\sigma$ (light blue) and 2$\sigma$ level (dark blue) contour regions spanning in the parameter space of electron neutrino anapole moment vs weak mixing angle, from a simulated XENONnT experiment with 60 tonne $\times$ source run, with a ${}^{51}$ Cr source, and considering statistical uncertainties only. It can be seen that the electron neutrino anapole moment could be tested at the 1$\sigma$ level in this experimental set up. We consider a range of values of the weak mixing angle spanning well beyond the current uncertainty range obtained from low-energy reactor neutrino experiments \cite{Canas:2016vxp}. It is clear that the SM expected value of the neutrino anapole moment is detectable at the 1$\sigma$ level in our experimental set-up, despite the uncertainty on the weak mixing angle. This statement relies on an uncertainty on the detector mass being smaller than 2.75$\%$, and on the activity of the Chromium source below 1$\%$, which is true for current liquid xenon detectors and ${}^{51}$Cr sources developed \cite{XENON:2023cxc, Danshin:2022wiz}.

For a sufficiently massive detector ($\gtrsim 60$ tonnes), a single source may yield a detectable event rate during a single source run. First proposals from the XLZD collaboration expect to reach 60 to 80 tonnes of detector mass, so in principle it would only be needed to replenish the source once to reach 1-2 $\sigma$ sensitivity to the neutrino anapole moment. Another possibility to pursue this science goal would be to repurpose the current XENONnT detector to search for the anapole moment of neutrinos, once the XLZD detector is operating. In that case, since the XENONnT detector has a mass of only 5.9 tonnes, it would be required to replenish the ${}^{51}$Cr source about $\sim$ 10 times before reaching the desired sensitivity to the neutrino anapole moment.

From Figure \ref{fig:limit_anapole_51Cr-electrons}, it can be noticed that our proposal significantly improves over the previously strongest bound as reported in the PDG \cite{ParticleDataGroup:2024cfk}, which stems from neutrino-electron scattering at the reactor neutrino experiment TEXONO \cite{TEXONO:2009knm}, using a CsI(Tl) scintillating crystal detector. Slightly stronger constraints on the electron neutrino charge radius have recently been derived from coherent elastic neutrino-nucleus scatterings at COHERENT \cite{AtzoriCorona:2022qrf} on a CsI detector. Prospects to set stronger bounds than TEXONO have also been derived for the LHC Forward Physics Facility \cite{MammenAbraham:2023psg}.

It should be noted that our proposal allows to go beyond the 1-2$\sigma$ detection level if a larger exposure and smaller systematical uncertainty (mainly from the fiducial mass of the detector) were achieved in the future. For instance, a reduction in the uncertainty of the detector mass to 1$\%$, and an exposure of 300 tonnes $\times$ source run would allow us to achieve sensitivity at the 3$\sigma$ level for values of the weak mixing angle greater than sin$^2 \theta_W \gtrsim 0.23$. We therefore encourage our experimental colleagues to further reduce the systematic uncertainty of the detector mass, which also plays an important role in the sensitivity prospects to the neutrino anapole moment, together with the total exposure.

\section{Sensitivity to the combination of electron and muon neutrino anapole moments}
In the following, we will work with an experimental set up consisting in a large exposure to the solar neutrino flux, of $10$ times larger than the exposure to a ${}^{51}$Cr source, and derive projected combined constraints on the electron neutrino anapole moment (present in both the solar and Chromium fluxes) and the muon neutrino anapole moment (present in the solar fluxes only). This is a plausible scenario if the Chromium source is operated during a source run of two months, but the experiment continues to run for other $\sim$ 20 months being sensitive to the solar neutrino flux, not too far from current liquid xenon timing exposures.

In this case, we will minimize of $\chi^2$ function over the range sin$^2 \theta_{W} = 0.22-0.25$, beyond the uncertainty range from low-energy neutrino experiments \cite{Canas:2016vxp}, so our choice is conservative. In the right panel of Figure \ref{fig:anapole_mue}, we show projected constraints in the parameter space of electron neutrino anapole moment vs muon neutrino anapole moment. We find that the SM prediction for the electron neutrino anapole moment can be probed in this case at 1$\sigma$, while the expected values of the muon neutrino anapole moment are off from the $2\sigma$ contours by more than an order of magnitude. The combination of the Chromium source and the solar neutrino flux with an exposure 19.5 times larger allows to increase the sensitivity to the electron neutrino anapole moment by $20\%$.

We also show in the right panel of Figure \ref{fig:anapole_mue} individual constraints from ${}^{51}$Cr and solar neutrinos on the parameter space spanned by the weak mixing angle vs electron neutrino anapole moment, after marginalizing over the muon neutrino anapole moment in the range $a_{\mu \mu}=[10^{-32}-10^{-34}]$ cm$^2$. Figure \ref{fig:anapole_mue} highlights the enhancement obtained from the ${}^{51}$Cr source even when increasing the exposure to solar neutrinos by a factor of 19.5. The improvement of the Chromium source arises mainly since it is composed of purely electron neutrinos, while the solar neutrino flux is distributed among all flavors.

\begin{figure*}[t!]
\includegraphics[width=0.48\linewidth]{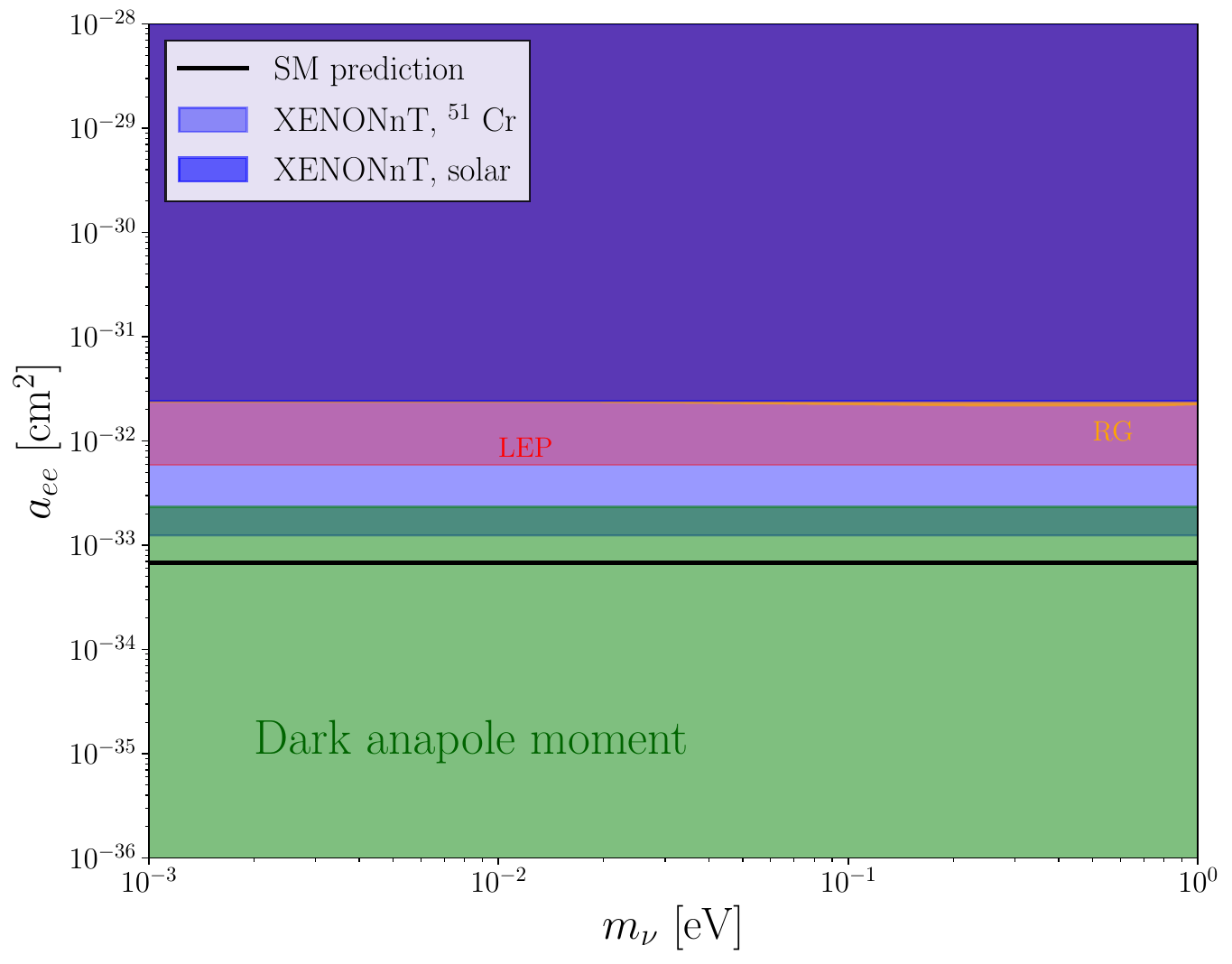}
\includegraphics[width=0.51\linewidth]{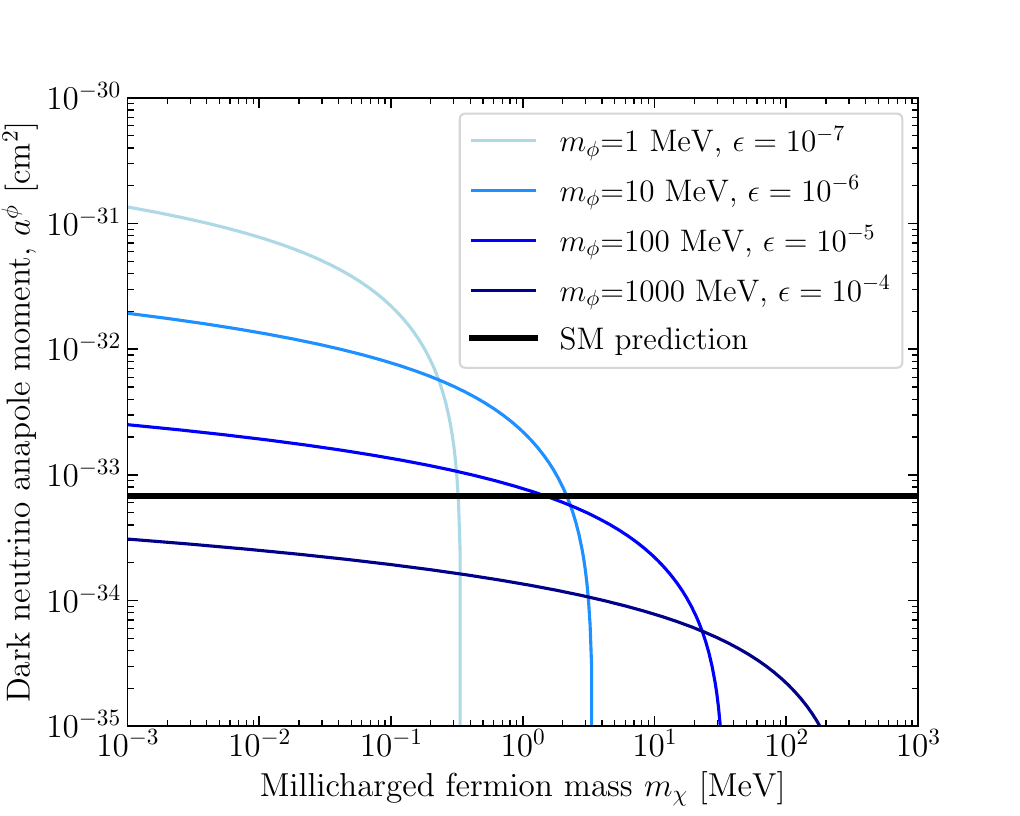}
    \caption{\textit{Left plot:} Predicted BSM values of the electron neutrino diagonal dark anapole moment in the parameter space considered in this work (shaded green), confronted with the SM expectation (solid black). The predicted values are allowed by complementary constraints on millicharged particles, see \textit{E.g} \cite{Vogel:2013raa}. For comparison, we show current constraints from solar neutrinos at XENONnT obtained in this work (fixing sin$^2 \theta_{W}=$0.2312), projected constraints from a ${}^{51}$Cr source and same exposure as currently, and complementary constraints from LEP and Red Giants \cite{Chu:2019rok}. \textit{Right panel}: Values of the dark neutrino anapole moment as a function of the millicharged fermion mass, compared to the SM prediction, for various values of the dark scalar mass and kinetic mixing. Here we take $Q=1$ and $(|c_L|^2 - |c_R|^2) = 1$ for concreteness.}
    \label{fig:anapole_BSM}
\end{figure*}
Finally, we discuss the possibility to constrain new physics in the neutrino sector via a future measurement of the neutrino anapole moment.
There are strong constraints on new electrically charged GeV scale particles from colliders, beam dump, indirect and direct detection experiments restricting such new particles to be very heavy. Neutrinos may be indirectly charged instead, evading some of these constraints.
For example, neutrinos may be coupled to new particles in the dark sector charged under a new $U(1)^{\prime}$ symmetry, whose corresponding gauge boson mixes kinetically with the SM photon (see \textit{E.g} \cite{Lindner:2017uvt, Jana:2024iig} for similar ideas).
Even though the neutrinos are not charged under this new symmetry, they may obtain dark moments at one-loop, and effective moments due to the kinetic mixing with the SM photon. This would naturally induce neutrino-electron interactions via an effective anapole moment.
If there are new scalars in the dark sector $\phi$, Dirac fermions $\chi$ and vector bosons $V$ charged under the dark $U(1)^{\prime}$ in units of $e$, and they couple to neutrinos, these give rise to dark electromagnetic moments of the neutrinos at the one-loop level. The light or massless gauge boson $A'$ of the $U(1)^{\prime}$ mixes with the SM photon, which gives a portal for the dark electromagnetic interaction to the visible sector. The strength is determined by the kinetic mixing $\epsilon$ via 
\begin{equation}
\mathcal{L} = \frac{\epsilon}{2} F'_{\mu\nu} F^{\mu\nu}
\end{equation}
where $F'_{\mu\nu}$ and $F^{\mu\nu}$ are the field strengths of the $U(1)^{\prime}$ and $U(1)$ gauge fields $A'$ and $A$ respectively. The interactions between neutrinos and the millicharged dark sector particles can be described by the Lagrangians 
\begin{equation}
\mathcal{L}_{\phi} = \bar \nu_{\alpha}(c^{\alpha}_{L}P_L+c^{\alpha}_{R}P_R) \phi^* \chi_{\alpha}
\end{equation}
and
\begin{equation}
\mathcal{L}_{\rm{V}}=  \bar \nu_{\alpha} \gamma^{\mu} (v^{\alpha}_{L}P_L+v^{\alpha}_{R}P_R) V_\mu \chi_{\alpha}+ \bar\nu_{\alpha} (g^{\alpha}_{R}P_R + g^{\alpha}_{R}P_R) G \chi_{\alpha}
\end{equation}

where $\chi_{\alpha}$ is a Dirac fermion, $V_\mu$ is a charged massive vector boson, $\phi$ a complex scalar field, and $G$ the longitudinal Goldstone polarization of $V_{\mu}$. These couplings lead at the one-loop order to an interaction with the dark photon $A^{\prime}$.  $c$, $g$ and $v$ are the couplings of the neutrinos to the scalar and vector particles, respectively. We note that the anapole moment can take positive or negative values depending on the relative strength of the left and right handed couplings.

Up to first order in the neutrino mass, the BSM contribution to the diagonal anapole moment of (Majorana) neutrinos due to new BSM particles in the loop has been calculated in a variety of works, \textit{E.g} \cite{Shrock:1982sc, Lucio:1984jn, Cabral-Rosetti:2002zyl, Lee:1977tib, Ibarra:2022nzm}. Useful compact expressions can be found in \cite{Ibarra:2022nzm}, whose normalization can be adjusted to account for the dark sector particles that are millicharged. For example, the anapole moment induced by a new millicharged fermion $\chi$ and scalar $\phi$, in a U(1)$^{\prime}$ extension of the SM with kinetic mixing $\epsilon$, reads
\begin{equation}
a^\phi \simeq  \frac{\epsilon \, e \, Q}{96 \pi^2 m_{\nu}^2}\left[\left|c_L\right|^2-\left|c_R\right|^2\right] \frac{3\left(\eta^2-\bar{\eta}^2\right)+\left(2 \eta^2+\bar{\eta}^2\right) \log \frac{\bar{\eta}^2}{\eta^2}}{\left(\eta^2-\bar{\eta}^2\right)^2}
\end{equation}
where we have defined the following ratios: 
\begin{equation}
\eta=\frac{m_{\chi}}{m_{\nu}} \, , \, \bar{\eta}= \frac{m_{\phi}}{m_{\nu}},
\end{equation}
which control the strength of the anapole moment. The dark anapole moment is enhanced when these ratios become small. We show an example of this in the right panel of Figure \ref{fig:anapole_BSM}, displaying the value of the neutrino anapole moment as a function of the dark millicharged mass, and for various values of the dark scalar mass.

We consider values of the dark sector particle masses at the MeV scale, and couplings and kinetic mixings allowed by current laboratory, astrophysical and cosmological constraints \cite{Davidson:2000hf, Vogel:2013raa}. In the left panel of Figure \ref{fig:anapole_BSM}, we show in shaded green color the range of values of the anapole moment of neutrinos obtained in this set-up, compared to the SM expectation. As can be appreciated, the neutrino anapole moment can be enhanced or suppressed with respect to the SM contribution, depending on the choice of parameters considered. It can be appreciated that our proposal to use a ${}^{51}$Cr source at XENONnT with a exposure of 60 tonne $\times$ source run allows to probe allowed parameter space of dark anapole moments. If no excess were detected, this search would set novel leading constraints on neutrinophilic light dark $U(1)$ sectors. Furthermore, the ${}^{51}$Cr search improves over complementary probes of dark anapole moments from electron-positron collisions at LEP, red giant anomalous cooling processes \cite{Chu:2019rok}, and current constraints from solar neutrino-electron scattering at XENONnT, which we derive in this work using the data from \cite{XENON:2022ltv}. It should be noted that the constraint from LEP only applies if the particles generating the
electromagnetic moment are sufficiently heavy compared to the center of mass energy of the experiment, at the GeV scale. Thus, for lighter masses of the loop particles as the ones discussed in the model proposed in this work, the constraint is not applicable.
Furthermore, in the left panel of Figure \ref{fig:anapole_BSM}, we show predictions on the dark anapole moment of neutrinos induced by a light dark sector, and various choices of parameters. Here it can be clearly appreciated that the anapole moment is enhanced when the mass of the loop-involved particles $m_{\phi}$ and $m_{\chi}$ is small.

\section{Conclusions}

We have shown that placing a ${}^{51}$Cr source nearby liquid xenon detectors may allow to detect at the 1$\sigma$ level the electron neutrino anapole moment predicted in the SM. The exposure required to achieve 1$\sigma$ sensitivity to the SM prediction of the neutrino anapole moment is about 60 tonne $\times$ source run, equivalent to 10 tonne $\times$ years, within reach of near future liquid xenon experiments such as XLZD.  Combining the radioactive source with a 19.5 times larger exposure to the solar neutrino flux only could allow to increase the sensitivity closer to the 2$\sigma$ level. Other mobile neutrino sources such as modular reactors may give an even larger enhancement on the neutrino flux, and the corresponding required exposure would be smaller. We leave this task for future investigation.

Furthermore, we have discussed how a future measurement of the anapole moment will allow to constrain new physics in the neutrino sector. In particular, we have discussed that active neutrinos can acquire a dark anapole moment at one-loop, and the kinetic mixing of the dark ultralight or massless gauge boson and the SM photon generates an effective anapole moment for the neutrinos. We have shown that new particles in the loop can enhance or suppress the anapole moment with respect to the SM contribution, which would allow to constrain models with light millicharged particles, which also couple to neutrinos.

\section*{Acknowledgments}

We are grateful to Ian M. Shoemaker and Garv Chauhan for useful discussions. The work was supported by the U.S. Department of Energy Office of Science under award number DE-SC0020262.

\bibliography{References}

\clearpage
\onecolumngrid

\end{document}